\documentclass[journal]{IEEEtran}

\usepackage{cite}
\usepackage{amsmath,amssymb,amsfonts}
\usepackage{algorithmic}
\usepackage{graphicx}
\usepackage{textcomp}

\usepackage{
amsmath,epsfig}

\usepackage{amsmath,epsfig,amsfonts,amssymb,graphics,theorem,calc,url,bm}
\usepackage{array}
\usepackage{calc}
\usepackage{verbatim}

\usepackage{amsmath}
\usepackage{amsfonts}
\usepackage{mathrsfs}
\usepackage{pifont}
\usepackage{amssymb}
\usepackage{verbatim}
\usepackage{upgreek}
\usepackage{color}
\usepackage{graphicx}
\usepackage{algorithmic}
\usepackage{algorithm}
\usepackage{epsfig}

\usepackage{fancybox}

\usepackage{amsfonts}
\usepackage{mathrsfs}
\usepackage{pifont}

\usepackage{verbatim}
\usepackage{upgreek}
\usepackage{color}

\usepackage{algorithmic}
\usepackage{algorithm}

\usepackage{fancybox}

\newcommand{\bzero}{\mbox{\boldmath{$0$}}}

\newcommand{\ba}{\mbox{\boldmath{$a$}}}

\newcommand{\bd}{\mbox{\boldmath{$d$}}}

\newcommand{\ii}{\mbox{\boldmath $i$}}

\newcommand{\bn}{\mbox{\boldmath{$n$}}}

\newcommand{\bQ}{\mbox{\boldmath{$Q$}}}
\newcommand{\bq}{\mbox{\boldmath{$q$}}}
\newcommand{\bR}{\mbox{\boldmath{$R$}}}

\newcommand{\bS}{\mbox{\boldmath{$S$}}}

\newcommand{\bW}{\mbox{\boldmath{$W$}}}
\newcommand{\bw}{\mbox{\boldmath{$w$}}}
\newcommand{\bX}{\mbox{\boldmath{$X$}}}
\newcommand{\bx}{\mbox{\boldmath{$x$}}}
\newcommand{\bY}{\mbox{\boldmath{$Y$}}}
\newcommand{\by}{\mbox{\boldmath{$y$}}}
\newcommand{\bZ}{\mbox{\boldmath{$Z$}}}

\newcommand{\bSigma}{\mbox{\boldmath{$\Sigma$}}}

\newcommand{\tr}{\mbox{\rm tr}\, }

\newcommand{\ex}{{\bf\sf E}}

\newcommand{\prob}{{\bf\sf Pr}}

\newtheorem{theorem}{Theorem}[section]

\newtheorem{lemma}[theorem]{Lemma}
\newtheorem{proposition}[theorem]{Proposition}

\begin{document}
%
\title{Robust Adaptive Beamforming Maximizing the Worst-Case SINR over Distributional Uncertainty Sets for Random INC Matrix and Signal Steering Vector}
%
%
%

\author{
Yongwei Huang, $\quad$ Wenzheng Yang, $\quad$ Sergiy A. Vorobyov
\thanks{Y. Huang and W. Yang are with School of Information Engineering, Guangdong University of Technology, University Town, Guangzhou, Guangdong, 510006, China. Emails: ywhuang@gdut.edu.cn, 2112003058@mail2.gdut.edu.cn.}%
\thanks{S. A. Vorobyov is with Department of Signal Processing and Acoustics, School of Electrical Engineering, Aalto University, Konemiehentie 2, 02150 Espoo, Finland. Email: svor@ieee.org.}
}

%

%



\maketitle

\begin{abstract}
The robust adaptive beamforming (RAB) problem is considered via the worst-case signal-to-interference-plus-noise ratio (SINR) maximization over distributional uncertainty sets for the random interference-plus-noise covariance (INC) matrix and desired signal steering vector. The distributional uncertainty set of the INC matrix accounts for the support and the positive semidefinite (PSD) mean of the distribution, and a similarity constraint on the mean. The distributional uncertainty set for the steering vector consists of the constraints on the known first- and second-order moments. The RAB problem is formulated as a minimization of the worst-case expected value of the  SINR denominator achieved by any distribution, subject to the expected value of the numerator being greater than or equal to one for each distribution. Resorting to the strong duality of linear conic programming, such a RAB problem is rewritten as a quadratic matrix inequality problem. It is then tackled by iteratively solving a sequence of linear matrix inequality relaxation problems with the penalty term on the rank-one PSD matrix constraint. To validate the results, simulation examples are presented, and they demonstrate the improved performance of the proposed robust beamformer in terms of the array output SINR.
\end{abstract}

\begin{IEEEkeywords}
Robust adaptive beamforming (RAB), distributionally robust optimization, strong duality, quadratic matrix inequality, rank-one solutions, interference-plus-noise covariance (INC) matrix.
\end{IEEEkeywords}

%
\IEEEpeerreviewmaketitle

\section{Introduction}

Robust adaptive beamforming (RAB) techniques are efficient for substantially enhancing the array (beamformer) output signal-to-interference-plus-noise ratio (SINR) and other array performance metrics and cope with mismatchs between the inaccurately known presumed and actual information about the source, propagation media and antenna array itself \cite{Stoica-Li-book-beamforming, GSSBO09, Voro13survey}. The latter means that the beamforming vector should always work well against any mismatch.

There is a large number of existing RAB designs (see, e.g., \cite{GSSBO09,Voro13survey} and references therein). Among the existing designs, the worst-case SINR maximization-based designs are especially popular. In \cite{VGL03tsp}, an optimal RAB design problem (in terms of the worst-case SINR maximization) is shown to be equivalent to a second-order cone programming (SOCP) problem when the mismatch between the actual desired signal steering vector and the  presumed steering vector is modeled as a ball centered at the presumed steering vector. It is also shown in \cite{Kim08} that the worst-case SINR maximization problem can be turned into a convex optimization problem, provided that the uncertainty sets for the interference-plus-noise covariance (INC) matrix and the signal steering vector are convex. In particular, the convex optimization problem is shown to be a semidefinite programming (SDP) problem as long as the uncertainty sets can be described by linear matrix inequalities (LMIs). In \cite{HFVL-CAMSAP2019-1}, the worst-case SINR maximization problem is reformulated into a quadratic matrix inequality (QMI) problem whenever the uncertainty set of the desired signal steering vector is a nonconvex set (e.g., the intersection of a ball centered at a nonzero point and a sphere centered at the origin) Moreover, a convex method of solving the QMI problem is proposed in \cite{HFVL-CAMSAP2019-1}.

In addition to the deterministic assumption about the mismatches, the case when the true signal steering vector (or the steering vector mismatch) is modeled as a random vector is also considered in \cite{Besson-tsp2004, Besson-tsp2005}. The probability chance constrained RAB problem (in terms of the SINR maximization) is shown in \cite{Voro-tsp08} to be tightly approximated by an SOCP problem in both cases of circularly symmetric Gaussian and worst-case distributions of the steering vector mismatch. When the distribution set for the signal steering vector contains the probability measure constraint and an ellipsoidal constraint of the first-order moment, the RAB problem (via the worst-case SINR maximization over the distribution set) in \cite{Zhang-icassp2015} is turned into a convex tractable form via the $S$-lemma \cite{Boyd}, if the support set is equal to the union of a finite number of sets, each of which is defined by a quadratic constraint. Finally, resorting to the linear conic reformulation of a distributionally robust chance constraint \cite{Zymler-2011MPA}, the distributionally robust beamforming problem is recast into an SDP problem in \cite{LiBin-SPL}, where the distributional robustness is understood as in \cite{Voro-tsp08}, i.e., finding the worst-case distribution in a set of all the possible distributions.

In this paper, we study a RAB problem based on distributionally robust optimization (DRO) (see, e.g., \cite{Delage-book}). The objective is to minimize the worst-case expected value of the interference-plus-noise power (the denominator of the SINR) achieved by any possible distribution for the random INC matrix in a distribution set. In addition, the constraint makes sure that the expected value of the useful signal power (the numerator of the SINR) is greater than or equal to one for all distributions of the random steering vector in another distribution set. Unlike the existing works (handling the random signal steering vector only while replacing the INC matrix by the sampling covariance matrix), the DRO model deals with the distributional uncertainty sets for both the INC matrix and the signal steering vector. Using the strong duality of linear conic programming, the DRO-based RAB problem is then reformulated into a QMI problem with respect to the beamforming vector, when the distribution set in the objective of the RAB problem consists of the probability measure constraint, the positive semidefinite (PSD) constraint, and a ball constraint on the first-order moment. The distribution set in the constraint of the RAB problem includes the constraints that are defined by the known first- and second-order moments. Then the QMI problem is solved via an LMI relaxation, and a rank-one solution for the LMI relaxation problem is obtained by an iterative procedure using the fact that the trace of a nonzero PSD matrix is equal to the Frobenius norm of the matrix if and only if it is of rank one. At each iteration, an LMI problem with a penalty term on the rank-one constraint in the objective is solved. Illustrative numerical examples are presented to show that the proposed method is better in terms of the array output SINR than two existing competitive beamforming techniques.

\section{Signal Model and Problem Formulation}

The array output signal is expressed as
\begin{equation}\label{array-output}
x(t)=\bw^H \by(t)
\end{equation}
where $\bw\in\mathbb{C}^N$ is the complex-valued beamforming
vector, $\by(t)\in\mathbb{C}^N$ is the snapshot vector of array observations, $N$ is the number
of antenna elements of the array, and $(\cdot)^H$ is the conjugate transpose. In the case of point signal source, the observation vector is given by
\begin{equation}\label{signal-observations}
\by(t)=s(t) \ba +\ii(t)+\bn(t)
\end{equation}
where $s(t)\ba$, $\ii(t)$, and $\bn(t)$ are the statistically
independent components of the signal of interest (SOI), interference, and
noise, respectively. In \eqref{signal-observations}, $s(t)$ is the SOI waveform and $\ba$ is its steering vector. For signal model \eqref{signal-observations}, the output SINR of the array is written as
\begin{equation}\label{SINR}
\mbox{SINR}=\frac{\sigma_s^2\bw^H\ba\ba^H\bw}{\bw^H\bR_{i+n}\bw}
\end{equation}
where $\sigma_s^2$ is the SOI power/variance and  $\bR_{i+n}\triangleq\ex\{ ( \ii (t) +\bn(t))(\ii(t) +\bn(t))^H\}$ is the INC matrix. Here $\ex\{\cdot\}$ stands for the statistical expectation. Thus, an optimal beamforming vector can be obtained by maximizing the SINR, which is equivalent to the following problem:
\begin{equation}\label{opt-beamforming-vector-nonrobust}
\underset{\bw}{\sf{minimize}}~\bw^H\bR_{i+n}\bw~~ {\sf{subject\;to}}~ |\bw^H\ba|^2\ge1.
\end{equation}

In practical scenarios, the INC matrix $\bR_{i+n}$ often is not available and  the true steering vector $\ba$ is also not known exactly. Considering both $\bR_{i+n}$ and $\ba$ as random variables (cf. \cite[Chapter 8]{Delage-book}), we formulate the DRO-based RAB problem as
\begin{equation}\label{rob-beamf-DRO}
\begin{array}[c]{cl}
\underset{\bw}{\sf{minimize}} &\begin{array}[c]{c} \underset{G_1\in{\cal D}_1}{\sf{max}}\ex_{G_1}\{\bw^H\bR_{i+n}\bw\}
\end{array}\\
\sf{subject\;to}               & \underset{G_2\in{\cal D}_2}{\sf{min}}\ex_{G_2}\{\bw^H\ba\ba^H\bw\}\ge1,
\end{array}
\end{equation}
where $\ex_{G_1}\{\cdot\}$ ($\ex_{G_2}\{\cdot\}$) denotes the statistical expectation under the distribution $G_1$ ($G_2$), and ${\cal D}_1$ and ${\cal D}_2$ are sets of distribution $G_1$ for random matrix $\bR_{i+n}\in{\cal H}^N$ (the set of all $N\times N$ Hermitian matrices)\footnote{The INC matrix $\bR_{i+n}$ can be treated as an $N^2$-dimension real-valued vector since there is an isomorphism between ${\cal H}^N$ and $\mathbb{R}^{N^2}$ (see, e.g., \cite{Andersen-book}).} and distribution $G_2$ for random vector $\ba$ (similarly, $\ba$ can be regarded as a $2N$-dimension real vector), respectively. Specifically, the distribution set ${\cal D}_1$ is defined as
\begin{equation}\label{distribution-set-D1}
\left\{G_1\in{\cal M}_1~\left|~\begin{array}{l}\prob_{G_1}\{\bR_{i+n}\in{\cal Z}_1\}=1\\ \ex_{G_1}\{\bR_{i+n}\}\succeq\bzero \\ \|\ex_{G_1}\{\bR_{i+n}\}-\bS_0\|_F\le\rho_1\end{array}\right.\right\}
\end{equation}
where ${\cal M}_1$ is the set of all probability measures on the measurable space $(\mathbb{R}^{N^2},{\cal B}_1)$, ${\cal B}_1$ is the Borel $\sigma$-algebra on $\mathbb{R}^{N^2}$, ${\cal Z}_1\subseteq\mathbb{R}^{N^2}$ is a Borel set. Also in \eqref{distribution-set-D1}, $\bS_0$ is the empirical mean of $\bR_{i+n}$, e.g., the sampling covariance matrix
\begin{equation}\label{sampling-matrix}
\hat\bR=\frac{1}{T}\sum_{t=1}^T\by(t)\by^H(t)
\end{equation}
where $T$ is the number of available training snapshots. Moreover, $\prob_{G_1}\{\cdot\}$ represents the probability of an event under the distribution $G_1$ and $\|\cdot\|_F$ is the Frobenius norm of a matrix.

To define ${\cal D}_2$, let us assume that the mean $\ba_0\in\mathbb{C}^N$ and the covariance matrix $\bSigma\succ\bzero$ of random vector $\ba$ under the true distribution $\bar G_2$ are known  (see \cite{Zymler-2011MPA}). The distribution set ${\cal D}_2$, which is a moment-based model, is then defined by
\begin{equation}\label{distribution-set-D2}
\left\{G_2\in{\cal M}_2~\left|~\begin{array}{l}\prob_{G_2}\{\ba\in{\cal Z}_2\}=1\\ \ex_{G_2}\{\ba\}=\ba_0 \\ \ex_{G_2}\{\ba\ba^H\}=\bSigma+\ba_0\ba_0^H\end{array}\right.\right\}
\end{equation}
where ${\cal M}_2$ is the set of all probability measures on the measurable space $(\mathbb{R}^{2N},{\cal B}_2)$, ${\cal B}_2$ is the Borel $\sigma$-algebra on $\mathbb{R}^{2N}$, and ${\cal Z}_2\subseteq\mathbb{R}^{2N}$ is a Borel set. 
 In other words, the set ${\cal D}_2$ includes all probability distributions on ${\cal Z}_2$ which have the same first- and second-order moments as $\bar G_2$.

Note that the difference between the DRO-based RAB problem \eqref{rob-beamf-DRO} and the distributionally robust beamforming problem considered in \cite{LiBin-SPL} are threefold. (i) In addition to the distributional uncertainty for the signal steering vector, we consider the distributional uncertainty for the random INC matrix. (ii) In the constraint of \eqref{rob-beamf-DRO}, we ensure that the expected value of the signal power $\bw^H\ba\ba^H\bw$ is greater than or equal to one for all $G_2\in{\cal D}_2$. In contrast, the distributionally robust chance constraint $\prob_P\{\Re(\bw^H\ba)\ge1\}\ge p$, $\forall P\in{\cal P}$ is considered in \cite{LiBin-SPL}. (iii) The distribution set in \eqref{distribution-set-D2} appears different from the distribution set ${\cal P}$ in \cite[Eq. (24)]{LiBin-SPL}.

\section{Reformulation for the DRO-based RAB Problem}
To tackle the DRO-based RAB problem \eqref{rob-beamf-DRO}, we first handle the inner maximization problem in the objective. Dropping the subscript of $\bR_{i+n}$ for notational simplicity, the inner problem is rewritten as
\begin{equation}\label{inner-max-objective}
\begin{array}[c]{cl}
\underset{G\in{\cal M}_1}{\sf{maximize}} &\begin{array}[c]{c} \int_{{\cal Z}_1}\bw^H\bR\bw\, dG_1(\bR) \end{array}\\
\sf{subject\;to}               & \int_{{\cal Z}_1}dG_1(\bR)=1\\
                               & \int_{{\cal Z}_1}\bR\, dG_1(\bR)\succeq\bzero\\
                               & \left\|\int_{{\cal Z}_1}\bR\, dG_1(\bR)-\bS_0\right\|_F\le\rho_1.
\end{array}
\end{equation}

The following proposition on the dual problem for \eqref{inner-max-objective} is in order.

\begin{proposition}\label{dual-inner-max-objective-prop}
The dual problem for \eqref{inner-max-objective} is cast as
\begin{equation}\label{dual-inner-max-objective}
\begin{array}[c]{cl}
\underset{}{\sf{minimize}} &\begin{array}[c]{c} \rho_1\|\bX\|_F+\delta_{{\cal Z}_1}(\bw\bw^H+\bX+\bY)-\tr(\bS_0\bX) \end{array}\\
\sf{subject\;to}               & \bX\in{\cal H}^N,\,\bY\succeq\bzero(\in{\cal H}_+^N),
\end{array}
\end{equation}
where $\delta_{{\cal Z}_1}(\cdot)$ stands for the support function of ${\cal Z}_1$ (see, e.g., \cite[Section 2.4]{beck-book2}), and ${\cal H}_+^N$ denotes the set of all $N\times N$ PSD Hermtian matrices. Further, the strong duality between \eqref{inner-max-objective} and \eqref{dual-inner-max-objective} holds.
\end{proposition}

The strong duality is due to Proposition 3.4 in \cite{Shapiro}, and a strictly feasible point for \eqref{inner-max-objective} can be constructed and the dual problem \eqref{dual-inner-max-objective} is always feasible. Observe that the support function $\delta_{{\cal Z}_2}(\cdot)$ is always convex, and hence, the dual problem \eqref{dual-inner-max-objective} is a convex problem. In particular, suppose that the set ${\cal Z}_1=\{\bR \in \mathbb{R}^{N^2}~|~\| \bR \|_F \le\rho_2\}$. Then the support function
\begin{equation}\label{support-fcn-1}
\delta_{{\cal Z}_1}(\bw\bw^H+\bX+\bY)=\rho_2\|\bw\bw^H+\bX+\bY\|_F.
\end{equation}

Let us take a look now to the minimization problem in the constraint of \eqref{rob-beamf-DRO}. The optimization problem can be equivalently rewritten as
\begin{equation}\label{inner-min-constraint}
\begin{array}[c]{cl}
\underset{G_2\in{\cal M}_2}{\sf{minimize}} &\begin{array}[c]{c} \int_{{\cal Z}_2}\ba^H\bw\bw^H\ba\, dG_2(\ba) \end{array}\\
\sf{subject\;to}               & \int_{{\cal Z}_2}dG_2(\ba)=1\\
                               & \int_{{\cal Z}_2}\ba\, dG_2(\ba)=\ba_0\\
                               & \int_{{\cal Z}_2}\ba\ba^H\,dG_2(\ba)=\bSigma+\ba_0\ba_0^H.
\end{array}
\end{equation}
Similar to Lemma A.1 in \cite{Zymler-2011MPA}, we can obtain the dual problem for \eqref{inner-min-constraint}, which is summarized in the following proposition.

\begin{proposition}\label{dual-inner-min-constraint-prop}
The dual problem for \eqref{inner-min-constraint} is given by
\begin{equation}\label{dual-inner-min-constraint}
\begin{array}[c]{cl}
\underset{}{\sf{maximize}} &\begin{array}[c]{c} x+\Re(\ba_0^H\bx)+\tr(\bZ(\bSigma+\ba_0\ba_0^H)) \end{array}\\
\sf{subject\;to}               & \ba^H(\bw\bw^H-\bZ)\ba-\Re(\ba^H\bx)-x\ge0,\,\forall\ba\in{\cal Z}_2\\
                               & \bZ\in {\cal H}^N,\,\bx\in\mathbb{C}^N,\,x\in\mathbb{R},
\end{array}
\end{equation}
where $\tr( \cdot )$ denotes the trace of a square matrix and $\Re( \cdot )$ stands for the real part of a complex-valued argument. Besides, the strong duality between \eqref{inner-min-constraint} and \eqref{dual-inner-min-constraint} holds.
\end{proposition}
Particularly, whenever ${\cal Z}_2=\mathbb{R}^{2N}$ (see \cite{Zymler-2011MPA}), the semi-infinite constraint in \eqref{dual-inner-min-constraint} can be recast into
\begin{equation}\label{constraints-in-dual-inner-min-constraint}
\left[ \begin{array}{cc}
 \bw\bw^H-\bZ&-\frac{\bx}{2}\\ -\frac{\bx^H}{2} & -x
\end{array} \right] \succeq \bzero.
\end{equation}

It follows from \eqref{dual-inner-max-objective}, \eqref{support-fcn-1}, \eqref{dual-inner-min-constraint} and \eqref{constraints-in-dual-inner-min-constraint} that the original RAB problem \eqref{rob-beamf-DRO} can be equivalently transformed into
\begin{equation}\label{rob-beamf-DRO-reformulation}
\begin{array}[c]{cl}
\underset{}{\sf{minimize}} &\begin{array}[c]{c} \rho_1\|\bX\|_F+\rho_2\|\bw\bw^H+\bX+\bY\|_F-\tr(\bS_0\bX) \end{array}\\
\sf{subject\;to}               &x+\Re(\ba_0^H\bx)+\tr(\bZ(\bSigma+\ba_0\ba_0^H))\ge1 \\
                               &\left[\begin{array}{cc}  \bw\bw^H-\bZ&-\frac{\bx}{2}\\ -\frac{\bx^H}{2} & -x \end{array} \right]\succeq\bzero\\
                               &\bw,\bx\in\mathbb{C}^N,\bX,\bZ\in{\cal H}^N,\,\bY\succeq\bzero,x\in\mathbb{R}.
\end{array}
\end{equation}

In addition, if ${\cal Z}_2$ is defined by a quadratic constraint, namely
\begin{equation}\label{Z2-quadratic-constraint}
{\cal Z}_2=\left\{\left[\begin{array}{c}\Re(\ba)\\ \Im(\ba)\end{array}\right]\in\mathbb{R}^{2N}~|~\ba^H\bQ\ba+2\Re(\bq^H\ba)+q\le0\right\},
\end{equation}
(here $\Im( \cdot )$ stands for the imaginary part of a complex-valued argument) then by the $S$-lemma (see, e.g., \cite{Boyd}), the semi-infinite constraint in \eqref{dual-inner-min-constraint} is tantamount to the following LMI:
\begin{equation}\label{LMI-Z2-quadratic-constraint}
\left[\begin{array}{cc}  \bw\bw^H-\bZ&-\frac{\bx}{2}\\ -\frac{\bx^H}{2} & -x \end{array} \right]\succeq \lambda \left[\begin{array}{cc}  \bQ&\bq\\ \bq^H & q \end{array} \right],\,\lambda\le0,
\end{equation} as long as there is a complex vector $\bar\ba$ such that $\bar\ba^H\bQ\bar\ba+2\Re(\bq^H\bar\ba)+q<0$. Therefore, in this case, the original RAB problem  \eqref{rob-beamf-DRO} is equivalent to problem \eqref{rob-beamf-DRO-reformulation} with the second constraint replaced by \eqref{LMI-Z2-quadratic-constraint}.

\section{A Rank-One Solution Procedure for the LMI Relaxation Problem for the DRO-Based RAB}
Observe that the DRO-based RAB problem \eqref{rob-beamf-DRO-reformulation} has a rank-one matrix $\bw\bw^H$ term in both the objective and the second constraint. Therefore, $\bw\bw^H$ can be relaxed as $\bW\succeq\bzero$, and the following LMI problem can be solved instead of the original problem \eqref{rob-beamf-DRO-reformulation}:
\begin{equation}\label{rob-beamf-DRO-reformulation-LMI}
\begin{array}[c]{cl}
\underset{}{\sf{minimize}} &\begin{array}[c]{c} \rho_1\|\bX\|_F+\rho_2\|\bW+\bX+\bY\|_F-\tr(\bS_0\bX) \end{array}\\
\sf{subject\;to}               &x+\Re(\ba_0^H\bx)+\tr(\bZ(\bSigma+\ba_0\ba_0^H))\ge1 \\
                               &\left[\begin{array}{cc}  \bW-\bZ&-\frac{\bx}{2}\\ -\frac{\bx^H}{2} & -x \end{array} \right]\succeq\bzero\\
                               &\bx\in\mathbb{C}^N,\bX,\bZ\in{\cal H}^N,\,\bW\succeq\bzero,\,\bY\succeq\bzero,x\in\mathbb{R}.
\end{array}
\end{equation}
If the solution $\bW^\star$ for \eqref{rob-beamf-DRO-reformulation-LMI}  is of rank one, that is $\bW^\star=\bw^\star\bw^{\star H}$, then $\bw^\star$ is an optimal beamforming vector for \eqref{rob-beamf-DRO-reformulation} as well. Otherwise, we need to find another rank-one solution for  \eqref{rob-beamf-DRO-reformulation-LMI}. Toward this end, we pay attention to the following immediate fact.

\begin{lemma}\label{rk-1-sufficient-condition} Suppose $\bW\succeq\bzero$ and $\bW\ne\bzero$. If $\tr(\bW)=\|\bW\|_F$, then $\bW$ is of rank one.
\end{lemma}

Observe that the condition $\tr\bW=\|\bW\|_F$ can be rewritten as
\[
\tr\bW-\frac{\tr(\bW\bW)}{\|\bW\|_F}=0.
\]
Therefore, an iterative procedure for finding a rank-one solution for \eqref{rob-beamf-DRO-reformulation-LMI} can be built. In the $k$-th step of such a procedure, the following LMI problem with a penalty term on the rank-one solution constraint, is solved:
\begin{equation}\label{rob-beamf-DRO-reformulation-LMI-iterative}
\begin{array}[c]{cl}
\underset{}{\sf{min}} &\rho_1\|\bX\|_F+\rho_2\|\bW+\bX+\bY\|_F-\tr(\bS_0\bX) \\
                       &~~~~~~~~~~~~~~~~~~~~~~~~~~~~+\alpha(\tr\bW-\frac{\tr(\bW\bW_k)}{\|\bW_k\|_F})\\
\sf{s.t.}               &x+\Re(\ba_0^H\bx)+\tr(\bZ(\bSigma+\ba_0\ba_0^H))\ge1 \\
                               &\left[\begin{array}{cc}  \bW-\bZ&-\frac{\bx}{2}\\ -\frac{\bx^H}{2} & -x \end{array} \right]\succeq\bzero\\
                               &\bx\in\mathbb{C}^N,\bX,\bZ\in{\cal H}^N,\,\bW\succeq\bzero,\,\bY\succeq\bzero,x\in\mathbb{R}
\end{array}
\end{equation}
where the penalty parameter $\alpha>0$ is a prefix (it is a large number) and the initial point $\bW_0$ is set to the high-rank solution $\bW^\star$ for \eqref{rob-beamf-DRO-reformulation-LMI}. Thereby, the iterative procedure for finding a rank-one solution for \eqref{rob-beamf-DRO-reformulation-LMI} is summarized in Algorithm~\ref{alg-1}.

\begin{algorithm}\caption{Finding a Rank-One Solution for Problem \eqref{rob-beamf-DRO-reformulation-LMI}}\label{alg-1}
\begin{algorithmic}[1]
\REQUIRE  $\bS_0$, $\bSigma$, $\ba_0$, $\rho_1$, $\rho_2$, $\alpha$; 

\ENSURE A rank-one solution $\bw^\star\bw^{\star H}$ for problem \eqref{rob-beamf-DRO-reformulation-LMI}; 

\STATE solve \eqref{rob-beamf-DRO-reformulation-LMI}, returning $\bW^\star$;

\IF {$\bW^\star=\bw^\star\bw^{\star H}$ is of rank one}

\STATE output  $\bw^\star$, and terminate;

\ENDIF

\STATE set $k=0$; let $\bW_k$ be the optimal (high-rank) solution $\bW^\star$ for  \eqref{rob-beamf-DRO-reformulation-LMI};

\REPEAT

\STATE solve the LMI problem \eqref{rob-beamf-DRO-reformulation-LMI-iterative}, obtaining solution $\bW_{k+1}$;

\STATE $k:=k+1$;


\UNTIL{$\left|\tr\bW_k-\frac{\tr(\bW_k\bW_{k-1})}{\|\bW_{k-1}\|_F}\right|\le 10^{-6}$}


\STATE output $\bw^\star$ with $\bW_k=\bw^\star\bw^{\star H}$.

\end{algorithmic}
\end{algorithm}

Note that the terminating condition in step 9 implies that $\bW_k$ is a rank-one solution for \eqref{rob-beamf-DRO-reformulation-LMI-iterative}, since $\tr(\bW_k)\approx\frac{\tr(\bW_k\bW_{k-1})}{\|\bW_{k-1}\|_F}\approx\|\bW_{k}\|_F$.


\section{Numerical examples}
Consider a uniform linear array of $N=10$ omni-directional sensor elements spaced half a wavelength apart from each other, and the power of additive noise in each sensor is set to 0 dB. Suppose that there are two interfering sources with the same interference-to-noise ratio (INR) of 30~dB impinging upon the array from the angles $\theta_1=-5^\circ$ and $\theta_2=15^\circ$. The angular sector $\Theta$ of interest is $[0^\circ,10^\circ]$ and the actual arrival direction of the desired signal is $5^\circ$, while the presumed direction is $\theta_0=1^\circ$. In addition to the signal look direction mismatch, we take into consideration mismatch caused also by wavefront distortion in an inhomogeneous media \cite{KVH-2012tsp}. In other words, we assume that the desired signal steering vector is distorted by wave propagation effects in the way that independent-increment phase distortions are accumulated by the
components of the steering vector. The phase increments are assumed to be independent Gaussian variables each with zero mean and standard deviation $0.02$, and they are randomly generated and remain unaltered in each simulation run. Suppose that the desired signal is always present in the training data and the training sample size $T$ in \eqref{sampling-matrix} is preset to 100. All results are averaged over 200 simulation runs.

Three beamformers are used for comparison, and they are obtained as the solutions of the following three RAB problems: problem \eqref{rob-beamf-DRO-reformulation-LMI}, the distributional RAB problem in \cite[Eqs.~(27)-(31)]{LiBin-SPL}, and the distributional RAB problem in \cite[Problem~(26)]{Zhang-icassp2015}. The beamformers obtained by solving the three aforementioned RAB problems are called ``Proposed beamformer", ``LRST beamformer", and ``ZLGL beamformer", respectively. In RAB problem \eqref{rob-beamf-DRO-reformulation-LMI}, $\bS_0$ is set to the sampling covariance matrix $\hat\bR$ in \eqref{sampling-matrix} (it is different in each run), the mean is $\ba_0=\frac{1}{L}\sum_{l=1}^L\bd(\theta_l)$, and the covariance is $\bSigma=\frac{1}{L}\sum_{l=1}^L (\bd(\theta_l) - \ba_0) (\bd(\theta_l) - \ba_0)^H$, where $\bd(\theta_l)$ is the steering vector associated with $\theta_l$ that has the structure defined by the sensor array geometry, and $\{\theta_l\}$ is a set of realizations of the random variable following the uniform distribution over the angular sector $\Theta$ of interest. In \eqref{rob-beamf-DRO-reformulation-LMI}, $\rho_1$ and $\rho_2$ are set to $0.001\|\bS_0\|_F$ and $10^5$, respectively, while in \eqref{rob-beamf-DRO-reformulation-LMI-iterative}, $\alpha$ is set to $10^{5}$.  When solving the beamforming problem in \cite{LiBin-SPL}, we set $\hat\bR_y=\hat\bR$ (see \eqref{sampling-matrix}), $\tilde{\ba}=\bd(\theta_0)$, the threshold $p=0.9$, and compute $\sigma^2$ in the same way as in the first numerical example in \cite{LiBin-SPL} (namely, the Gaussian mixture model is considered). In the beamforming problem (26) of \cite{Zhang-icassp2015}, we set $\hat\bR_y=\hat\bR$, $\bd_l=[\Re(\bd(\theta_l))^T,\Im(\bd(\theta_l))^T]^T$ (where $\{\theta_l\}$ is another set of realizations of the uniform distribution over the angular sector $\Theta$), $\hat\ba=\frac{1}{L}\sum_{l=1}^L\bd_l$, $\hat\bSigma=\frac{1}{L}\sum_{l=1}^L(\bd_l-\hat\ba)(\bd_l-\hat\ba)^T$, and $\gamma=0.1N$. Also the support set ${\cal S}$ is equal to the union of three ellipsoids ($n=3$ in problem (26) of \cite{Zhang-icassp2015}), i.e., ${\cal S}={\cal E}_1(\hat\ba_1,\gamma_1\hat\bSigma_1)\cup{\cal E}_2(\hat\ba_2, \gamma_2 \hat\bSigma_2)\cup{\cal E}_3(\hat\ba_3,\gamma_3\hat\bSigma_3)$, where we set $\gamma_i=0.2N$, $i=1,2,3$, and the entries of $(\hat\ba_i,\bQ_i)$, $i=1,2,3$, are generated randomly from the standard Gaussian distribution with $\hat\bSigma_i=\bQ_i\bQ_i^T$. 

\begin{figure}[!h]
\centerline{\resizebox{.42\textwidth}{!}{\includegraphics{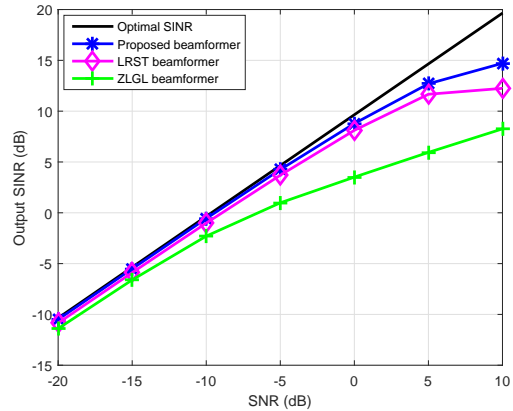}}
}
\vspace*{-.5\baselineskip}
\caption{Average array output SINR versus SNR with $T=100$.}
\label{fig-sinr-snr}
\vspace*{0\baselineskip}
\end{figure}

\begin{figure}[!h]
\centerline{\resizebox{.42\textwidth}{!}{\includegraphics{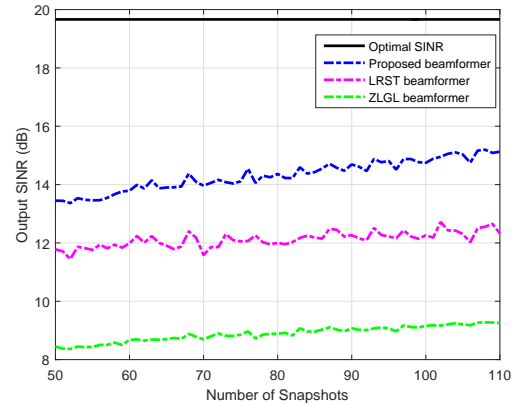}}
}
\vspace*{0.3\baselineskip}
\caption{Average beamformer output power versus number of snapshots with SNR equal to 10~dB.}
\label{fig-sinr-snap}
\vspace*{0\baselineskip}
\end{figure}

Fig.~\ref{fig-sinr-snr} displays the sensor array output SINR versus the  signal-to-noise ratio (SNR) for a single antenna when number of snapshots $T=100$. 
It can be seen that the output SINR obtained through \eqref{rob-beamf-DRO-reformulation-LMI} is higher than those obtained by the other two beamformers at the region of moderate SNRs. Fig.~\ref{fig-sinr-snap} demonstrates the beamformer output SINR versus the number of snapshots with SNR equal to 10 dB. In can be observed that our proposed beamformer yields better SINR than the other two beamformers.

\section{Conclusion}
We have considered the DRO-based RAB problem of maximizing the worst-case output SINR over the distributional uncertainty sets for the INC matrix and the signal steering vector. The equivalent problem of minimizing the worst-case expected value of the interference-plus-noise power over the distribution set for the INC matrix subject to the expected value of the signal power to be greater than or equal to one for all distributions in the distribution set for the signal steering vector has been formulated. Using the strong duality of linear conic programming, the DRO-based RAB problem has been equivalently transformed into a QMI problem, which has been solved by a sequence of LMI problems with a penalty term on the rank-one constraint in the objective function. The improved performance of the proposed DRO-based RAB has been demonstrated by simulations.

\end{document}